\documentclass[pdflatex,sn-mathphys-num]{sn-jnl}

\usepackage{graphicx}%
\usepackage{multirow}%
\usepackage{amsmath,amssymb,amsfonts}%
\usepackage{amsthm}%
\usepackage{mathrsfs}%
\usepackage[title]{appendix}%
\usepackage{xcolor}%
\usepackage{textcomp}%
\usepackage{manyfoot}%
\usepackage{booktabs}%
\usepackage{algorithm}%
\usepackage{algorithmicx}%
\usepackage{algpseudocode}%
\usepackage{listings}%

\usepackage{bm}
\usepackage{braket}

\begin{document}

\title[Localized Covariant Quantities Appear To Underlie Quantum Circuits]{Localized Covariant Quantities Appear To Underlie Quantum Circuits}

\author[1]{\fnm{Ken} \sur{Wharton}}

\author[2]{\fnm{Roderick} \sur{Sutherland}}

\author[1]{\fnm{Titus} \sur{Amza}}

\author[1,3]{\fnm{James} \sur{Saslow}}

\affil[1]{\orgdiv{Department of Physics and Astronomy}, \orgname{San Jos\'e State University}, \orgaddress{\street{One Washington Sq.}, \city{San Jos\'e}, \postcode{95192-0106}, \state{CA}, \country{USA}}}

\affil[2]{\orgdiv{Centre for Time}, \orgname{University of Sydney}, \orgaddress{\city{Sydney}, \state{NSW}, \country{Australia}}}

\affil[3]{\orgname{QuantWare}, \orgaddress{\city{Delft}, \country{Netherlands}}}

\abstract {Although entangled state vectors cannot be fully described in terms of variables localized in space and time, any given entanglement \textit{experiment} can be built from basic quantum circuit components with well-defined locations. We analyze such quantum circuits and present evidence that the local weak values comprise a covariant tensor associated with each individual qubit. Even if the state is massively entangled, these tensors do not evolve or collapse when other qubits are measured or pass through distant circuit elements. They can therefore be viewed from different reference frames without contradiction. Furthermore, their evolution through any circuit always obeys covariant dynamical rules. Weak values are subject to both past and future constraints, so the covariant quantities can only be determined by considering the entire circuit ``all-at-once'', as in action principles, incorporating the future measurement basis to avoid the standard no-go theorems. Because these results hold for a set of universal quantum gates, this work lends support to the claim that any quantum circuit can be assigned a realistic, lower-level description compatible with our understanding of classical spacetime.}

\maketitle

\section{Introduction}\label{sec1}

In the context of quantum entanglement experiments with eventual (strong) measurements, it may seem impossible to find a relativistically-covariant description of what is happening in spacetime \cite{frame1,frame2,frame3}. One obvious obstacle is that the conventional quantum description of an entangled wavefunction is a function on a multi-particle configuration space, $\psi(\bm{x}_1,\bm{x}_2,...,\bm{x}_N,t)$, while covariant quantities (``covariants'') are instead associated with events in ordinary spacetime. Even if one extracted spacetime-localized variables from the wavefunction (say, via a partial trace, or by adding hidden Bohmian particle positions), the dynamics of that description would seem to be strongly frame-dependent. In the simple quantum circuit shown in Figure 1, even with well-defined qubit worldlines, it is difficult to find any quantity associated with spacetime region $Q$. In particular, the entangling $\sqrt{SWAP}$ gate and the strong measurement would seem to cause inherent contradictions between the three reference frames identified by planes of simultaneity\footnote{Dismissing this concern as only relevant for relativistic quantum mechanics would miss the point that for a sufficiently large system, the relative velocities between these three frames might be extremely small.} in Figure 1.

\begin{figure}[htbp]
\begin{center}
\includegraphics[width=8cm]{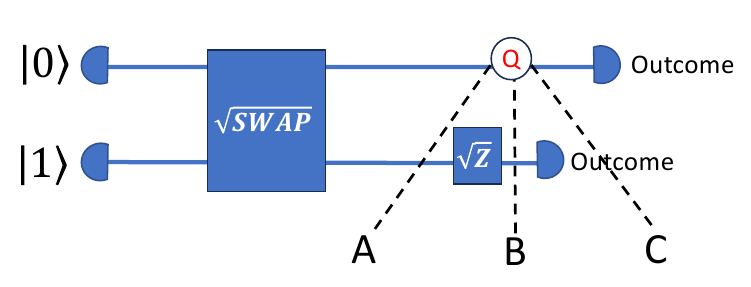}
\caption{A two-qubit quantum circuit which might be used in a test of Bell-inequality violations. Time runs to the right, with the qubits becoming maximally-entangled, then encountering strong measurements. The dotted lines represent simultaneity in different reference frames, each intersecting the same event at Q. In frames $A$ and $B$ the two qubits are described by different entangled states, because in frame $A$ the $\sqrt{Z}$ gate has not yet been encountered. In frame $C$ the qubits are described by separate, collapsed states due to the completed measurement on the lower qubit. The partial trace at $Q$ would be at the center of the Bloch ball in frames $A$ and $B$, but on the surface of the Bloch sphere in frame $C$.} 
\label{default}
\end{center}
\end{figure} 

Despite these apparent difficulties, this paper will identify relativistically-covariant quantities associated with individual qubits throughout such quantum circuits. Furthermore, these quantities will always evolve according to covariant equations. To the extent that any quantum system can be accurately modeled by a sufficiently large quantum circuit, this is a general result that seems likely to have important implications for a number of foundational questions. For instance, the long-standing disconnect between quantum theory and general relativity is (in part) due to the fact that the former is formulated on configuration space rather than on a spacetime manifold. If there are spacetime-localized covariants associated with arbitrary quantum systems, identifying all of them could be a necessary step towards resolving this fundamental mismatch.

More generally, focusing on the covariant properties of a given theory is one way to reformulate a nonlocal theory into a local one. For instance, in classical electromagnetism, the Coulomb-gauge potentials behave nonlocally, appearing to allow instantaneous control of distant potentials. This behavior disappears when attention is focused on the covariant Lorentz-gauge potentials.  Alternatively (and more analogous to what follows), one might identify the spacetime-local invariants, $E^2-c^2B^2$ and $\vec{E}\cdot\vec{B}$, and the covariant field tensor $F^{\mu\nu}$ from which these invariants are generated.  Drawing attention to the covariant field then naturally downplays any imagined importance of Coulomb-gauge potentials. In the same way, by focusing attention on the covariant aspects of quantum circuits, one can hope for a similar outcome. If successful, this would provide evidence against the ontological status of the configuration-space wavefunction, instead drawing attention to the underlying covariant quantities.

In the case of quantum circuits, this analysis cannot be quite as simple as electromagnetism; one also needs to resolve the relevant no-go theorems (primarily Bell's theorem \cite{bell2004}, but also Kochen-Specker \cite{KS} and Pusey-Barrett-Rudolph \cite{PBR}) by rejecting at least one of their assumptions. This challenge can be met by using \textit{complete} quantum circuits, where the covariant quantities are calculated using the entire circuit -- in particular, the final measurement basis and the eventual outcomes. In ordinary classical situations, learning about outcomes allows one to better reconstruct what has already happened, and there is little reason to think that quantum systems would be any different. However, the below analysis will require not merely the outcomes, but also the controllable aspects of the circuit itself. Using this information may seem more problematic, but it is also used in any path-integral calculation, where the entire experimental setup must be considered ``all-at-once''. The above-mentioned quantum no-go theorems each \textit{assume} that the future measurement settings should be irrelevant to facts in the past. Breaking this time-asymmetric assumption is therefore a natural way to restore a spacetime-localized description to quantum phenomena \cite{wharton2020}.

The well-known quantum circuit framework is ideal for exploring relativistic covariance in this manner. Quantum circuits are modular; every possible quantum circuit can be built from a few basic one- and two-qubit gates, and the behavior of these simple gates is easy to analyze.  An obvious spacetime structure is also evident from the circuit itself. The qubit ``wires’’ (the lines in Figure 1) can represent actual ``worldlines'' (say, paths of physically localized particles). The two-qubit gates indicate where interactions occur: where the worldlines are physically brought together. Furthermore, any complete quantum circuit has a pre-chosen measurement basis at its end, and this information is therefore available when implementing any ``all-at-once'' analysis of the full circuit.  

This last point assumes the standard ``block-universe'' framework from relativity, which is fitting given the goal of finding relativistic covariants. Most arguments against this all-at-once viewpoint take time to be ``flowing forward'' in some sense\footnote{Note that any ``flow'' viewpoint cannot utilize the dictionary-definition of \textit{flow}, which means motion through a spatial dimension.} and therefore would discount the importance of future events when analyzing the present instant. But even defining ``the present'' is a notoriously frame-dependent concept; if one does not allow any special foliation of spacetime, then a block universe view is arguably more appropriate. 

The primary analysis of these circuits will be done using a well-known technique that appears to probe local properties of a quantum system: post-selected weak values. These values have been independently derived in a number of different contexts \cite{roberts1978, AAV, sutherland1998}, and they can also be measured in the laboratory (by averaging many interactions) \cite{exp1,exp2,exp3,exp4,nature1}.  Post-selected weak values naturally incorporate information about the eventual measurement basis, making them ``future-input dependent'' as defined in \cite{wharton2020}. It is not common to use local weak values to analyze properties of entangled states (see Section 7 of \cite{sutherland2022} for an exception), but they are naturally suited to do precisely that.  Most of the central results in this paper will come from analyzing the behavior of the weak values as they pass through various quantum gates, and noting that they behave just like local properties. For instance, in the midst of a continuous two-qubit gate, one can see that these weak values evolve according to a simple, local, differential equation, even if the quantum state in question is completely entangled. 

As a counterpoint to this approach, Barandes has recently cautioned against any literal reading of weak values as physical properties \cite{barandes}. He writes, ``To claim ... that weak values have some specific interpretation, let alone an interpretation that makes sense at the level of a single system, would be an extraordinary assertion, and the burden should be on the claimants to provide a rigorous argument in favor of that view.'' This paper aims to meet this challenge, and we will return to a discussion of Barandes' arguments in the conclusion.

Taken together, this paper presents several strands of evidence that quantum systems seem to be associated with an underlying local, relativistic structure. The present results are mainly concerned with identifying the localized covariant quantities and dynamics in a quantum circuit framework, but they are also part of a larger research program that seeks to eliminate the state vectors from the fundamental description (relegating them to mere states of knowledge about the underlying covariants; see \cite{fuchs,spekkens,wharton2014})\footnote{This is but one potential use of these results. It should go without saying that many other foundational approaches might also find uses for the covariant quantities identified here.}. This larger goal cannot hope to be successful without first identifying the correct underlying ``ontology’’, and that is the focus of this paper. Although state vectors are used to calculate the covariants in the below analysis, some of our results hint at an alternative calculation order, in which these covariants can be calculated without state vectors. For readers interested in such an eventual reformulation, the present results should be viewed as a framework and motivation for this next stage of research. But our primary conclusion is that local weak values of quantum circuits comprise a covariant structure with covariant dynamics, and these novel results can speak for themselves.

\section{Post-selected Weak Values}\label{sec2}

\subsection{Overview and Definitions}

We will consider a general quantum circuit that begins with a preparation procedure at time $t_i$, determining the initial state vector $\ket{i}$. The known circuit elements allow one to calculate the state vector at the time of final measurement $t_f$ as ${U}[t_f-t_i]\ket{i}$, where ${U}[t_f-t_i]$ is notation used to summarize the unitary operations of the circuit from time $t_i$ to time $t_f$.  Of course, this calculated state vector is usually not what is actually observed at time $t_f$. Instead, some eigenstate $\ket{f}$ of the final measurement basis is inferred from the measurement outcome.  In standard quantum mechanics, this is the newly collapsed state.  For any given run of any given quantum circuit, $\ket{f}$ provides additional information as to what has actually happened between times $t_i$ and $t_f$.

If this previous claim seems surprising, consider that in any classical situation with some unknown details, learning about outcomes will generally provide information about events prior to those outcomes.  In a quantum context, this fact has been formally proven -- both theoretically \cite{AAV, dressel2015} and in the laboratory \cite{exp1,exp2,exp3,exp4} -- in the context of post-selected weak values.  Given the input, output, and structure of a quantum circuit, a weak value $W$ can be defined for any Hermitian operator ${A}$ at any intermediate time $t$ according to
\begin{equation}
\label{eq:weak}
    W[A](t) =\frac{\bra{f}{U}[t_f-t]\,{A}\,{U}[t-t_i]\ket{i}}{\braket{f|{U}[t_f-t_i]|i}}.
    \end{equation}
Experimental validations of this expression involve actual ``weak measurements'' of ${A}$, which must be repeated many times to bring the signal out of the noise, and then ``post-selected'' for the particular outcome $\ket{f}$ of interest.  When this procedure is performed, the averaged post-selected results match the predicted weak value in the limit of minimal coupling between the system and the weak measurement device.

It must be noted that the above expression is mathematically well-defined even if no ``weak measurement'' is actually carried out.  As Dressel puts it \cite{dressel2015}, ``one then interprets Eq. (\ref{eq:weak}) as the best estimate of the (unmeasured) average value of ${A}$ \dots given only $\ket{i}$,$\bra{f}$, and [${U}$]''.  This means there is no restriction on simultaneously considering the weak values of non-commuting observables. For a single qubit, the weak values for all three Pauli operators $\sigma_x$, $\sigma_y$ and $\sigma_z$ can be calculated together. Inserting these three operators into Eq. (1) yields the x, y and z components of a weak value vector:
\begin{equation}
\label{eq:sdef}
   \bm{w}=(W[\sigma_x],W[\sigma_y],W_[\sigma_z]). 
\end{equation}
As these three operators span all non-trivial $2\times 2$ Hermitian matrices, this vector $\bm{w}$ will be the focus of the below analysis.

In general, the weak values are complex numbers, so $\bm{w}$ can be decomposed into two 3-vectors, $\bm{w}=Re(\bm{w})+i Im(\bm{w})$, or else viewed as a single \textit{complex} 3-vector.  Weak values are not dependent on the global phase; any global phase adjustment to either $\ket{i}$ or $\ket{f}$ will appear in both the numerator and denominator of Eq. (\ref{eq:weak}) and will cancel.  The complex values of $\bm{w}$ are therefore not phase-dependent; there is an objective distinction between the vectors $Re(\bm{w})$ and $Im(\bm{w})$.

Note that the weak values are not restricted to the usual eigenvalues of the corresponding operator.  The values of $Re(W[A])$ can be much larger than the maximum eigenvalue of $A$ in cases where the probability of the actual outcome is small. Nevertheless, in the usual situation where the result of the final measurement is unknown and a weighted average is taken over the possible outcomes, the weak value $Re(W[A])(t)$ can then be shown to be exactly equal to the usual expectation value $\langle A\rangle (t)$. Hence the weak value tells us nothing new until we post-select a particular final outcome and look at the weak value for just that subset.

It is also possible to generate weak values associated with multiple qubits at once, because the operator $A$ in Eq. (\ref{eq:weak}) is not necessarily restricted to single-qubit operators.  But any such extension would defeat the purpose of this analysis, which is to find local covariants associated with each individual qubit in a quantum circuit.  For this reason the weak values of such multi-qubit operators will not be considered in this paper; any use of the term ``weak values'' will be assumed to mean \textit{local} weak values, only concerning the properties of single qubits.

\subsection{Single Qubit Example}\label{sec22}

Consider a qubit prepared in the state $\ket{i}=\ket{0}$ and then measured in the basis
\begin{eqnarray}
    \ket{f_+}&=&\cos \frac{\theta_0}{2} \ket{0} + \sin \frac{\theta_0}{2} \ket{1}\label{eq:eq3}\\
    \ket{f_-}&=&\sin \frac{\theta_0}{2} \ket{0} - \cos \frac{\theta_0}{2} \ket{1} \label{eq:eq4}
\end{eqnarray}
for some chosen value of $\theta_0<\pi$. (On the Bloch sphere, this assumes the measurement of $\ket{f_+}$ corresponds to some direction in the $+\hat{x}$ half of the $x-z$ plane.) This example is actually quite general, as there is always a coordinate rotation which could transform any single-qubit prepare-and-measure scenario into this one.  

Throughout this paper we will want to refer to such states as unit 3-vectors on the Bloch sphere\footnote{One way to find these vectors from their corresponding states is $\ket{f}\bra{f}=(I+\hat{\bm{f}}\cdot\bm{\sigma})/2$, where $\bm{\sigma}$ is the usual vector of Pauli matrices.}, in this case $\hat{\bm{f}}_+$ and $\hat{\bm{f}}_-$. Here the ``hat'' notation will be used for unit vectors corresponding to single-qubit states, so that $\hat{\bm{i}}$ represents the Bloch sphere 3-vector representation of $\ket{i}$, etc.

With this notation, the above example corresponds to preparing a state in the $\hat{\bm{i}}=\hat{\bm{z}}$ direction and then measuring it to be in either the $\hat{\bm{f}}_+=\sin\theta_0\hat{\bm{x}}+\cos\theta_0\hat{\bm{z}}$ direction for the outcome $\ket{f_+}$, or in the opposite direction, $\hat{\bm{f}}_-=-\hat{\bm{f}}_+$ for the outcome $\ket{f_-}$.  The Hamiltonian in this example is zero; there is no time evolution of the qubit between preparation and measurement.  It follows from Eq. (\ref{eq:weak}) that $\bm{w}$ will therefore be constant as well.

 The weak values of the qubit at every point between preparation and measurement can be easily calculated for either measured outcome, $\ket{f_\pm}$.  Referring back to the vector defined in Eq. (\ref{eq:sdef}), each component can be calculated from (\ref{eq:weak}) by setting $A$ to the corresponding Pauli matrix.  For $\ket{f_\pm}$ in Eqs. (\ref{eq:eq3}) and (\ref{eq:eq4}), the two possible forms for $\bm{w}$ are found to be
\begin{eqnarray}
    \bm{w}_+ &=& \tan(\theta_0/2)\hat{\bm{x}} + i\tan(\theta_0/2)\hat{\bm{y}} +  \hat{\bm{z}}  \\
    \bm{w}_- &=&  -\cot(\theta_0/2) \hat{\bm{x}} -i\cot(\theta_0/2)\hat{\bm{y}} + \hat{\bm{z}} 
\end{eqnarray}
These weak value vectors are complex, unnormalized, and unitless. If the qubit were a spin-1/2 particle, the only difference between these vectors $\bm{w}$ and the weak value vector of the spin-1/2 operators $(W[{S}_x],W[{S}_y],W[{S}_z])$ is that the latter vector would have an extra factor of $\hbar/2$.

Like $\hat{\bm{f}}_+$ and $\hat{\bm{f}}_-$, the real part of each weak value vector also lies in the $x-z$ plane, as shown in Figure 2. The real component of $\bm{w}$ bisects the angle between the initial and final states, and its length can also be found geometrically, as shown by the dashed lines in Figure 2.  A possible interpretation of this geometry will be discussed below.

\begin{figure}[htbp]
\begin{center}
\includegraphics[width=8cm]{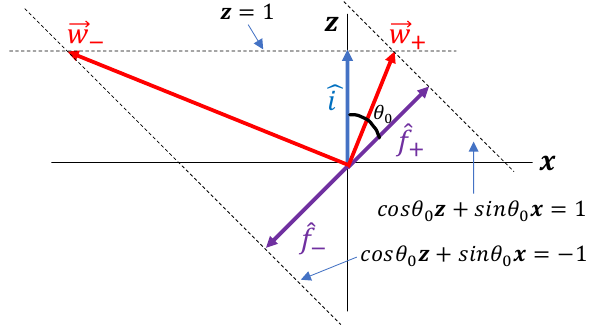}
\caption{The x-z plane of the Bloch sphere, showing the prepared state $\hat{\bm{i}}$, the two possible measured states $\hat{\bm{f}}_\pm$, and the real part of the two corresponding weak value vectors $\vec{\bm{w}}_\pm=Re(\bm{w}_\pm)$. These vectors lie at the intersection of two constraints corresponding to the initial preparation and final measurement (indicated by dashed lines).}
\end{center}
\end{figure}

Again, the magnitudes of the vectors are not normalized. Instead, one finds $[Re(\bm{w}_+)]^2=\sec^2(\theta_0/2)$ and $[Re(\bm{w}_-)]^2=\csc^2(\theta_0/2)$. Curiously, these square magnitudes are inversely proportional to the probabilities of the corresponding outcomes. Less surprisingly, the weighted average of $\bm{w}_+$ and $\bm{w}_-$ (using their Born-rule probabilities) is exactly $\hat{\bm{z}}$, with no imaginary part surviving. This average matches the preparation $\hat{\bm{i}}$.

Even though the vectors $Re[\bm{w}_\pm]$ both extend beyond the surface of the Bloch sphere, there is still a sort of ``hyperbolic normalization''. Treating the full $\bm{w}$ as a complex 3-vector, it is natural to define the complex scalar
\begin{eqnarray}
\label{sdots}
    \bm{w}\cdot\bm{w} &\equiv& [Re(\bm{w}) + iIm(\bm{w})]\cdot[Re(\bm{w}) + iIm(\bm{w})]\\
    &=& [Re(\bm{w})]^2 - [Im(\bm{w})]^2 + 2i Re(\bm{w})\cdot Im(\bm{w}).
\end{eqnarray}
Notice the minus sign on $[Im(\bm{w})]^2$; no complex conjugates have been taken.  Given this definition, one finds $\bm{w}\cdot\bm{w}=1$ (with a zero imaginary part) for both $\bm{w}_+$ and $\bm{w}_-$.

\subsection{Discussion} \label{sec23}

Weak values have no widely accepted interpretation, but Roberts \cite{roberts1978} originally proposed that his related quantities could represent ``part of objective reality'', some hidden feature of the actual quantum system.  This viewpoint has also been advocated and extended by Sutherland \cite{sutherland1998,sutherland2008,sutherland2017}.  

It is important to note a crucial distinction between these proposals and the so-called ``Two State Vector Formalism'' (TSVF) \cite{tsv}, in which the focus is on the (unlocalized) state vectors as elements of reality, as opposed to the (localized) weak values themselves.  Analysis of TSVF sometimes implies the reality of weak values, as something which the ``particles are allowed to possess'' \cite{aharonov2015}, and more recently has led to a similar ``weak value interpretation'' of TSVF itself \cite{waegell2023}.  But more often TSVF treats weak values as arising at the time of weak measurements, rather than as continuously existing properties. In that view, weak values are then seen as a consequence of the state vectors rather than an explanatory resource in their own right.

Even though the weak values are here calculated using the full state vectors, it is natural to ask whether one can make sense of the weak values on their own terms, without any reference to state vectors at all.  This is indeed possible for the special case of a single non-interacting qubit (from \ref{sec22}).  With $\bm{w}$ remaining constant from preparation to measurement, it becomes possible to compute the intermediate $Re[\bm{w}]$ using only the initial preparation and the final measurement result, as shown geometrically in Figure 2.  Then, $Im[\bm{w}]$ can be determined within a sign\footnote{Furthermore, the sign is fixed by certain handedness conditions at preparation and measurement; see \cite{preprint}.} using $\bm{w}\cdot\bm{w}=1$.

A physical interpretation of this weak value is also evident, at least for the case where the qubit is a spin-1/2 particle.  The initial preparation can be interpreted as choosing the result of an initial spin measurement in the $\hat{\bm{z}}$ direction, fixing the $z$-component of the weak value $\bm{w}\cdot\hat{\bm{z}}=1$.  No other component is measured at this moment, so there is no initial constraint on any other component of $\bm{w}$, or even its total magnitude.  The final measurement then constrains a \textit{different} component of $\bm{w}$ to be $\pm 1$ (either $\hat{\bm{f}}_+$ or $\hat{\bm{f}}_-$).  For each possible result, the smallest vector that conforms to both of these constraints, without changing between measurements, is precisely $Re(\bm{w}_\pm)$.  Such a hidden structure of a single spin-1/2 particle, as described by the weak value formalism, has previously been explored in Sec. 6 of \cite{sutherland2022}.  

While it is true that state vectors were used for calculating the probabilities of the two possible outcomes, even this use of the state vector can be easily discarded in this particular example. As noted above, the Born rule could be replaced with a global postulate that assigns probabilities directly to the weak values:
\begin{equation}
\label{eq:prob1}
    P(\bm{w}) \propto \frac{1}{|Re(\bm{w})|^2}.
\end{equation}
Therefore, since one knows the two possible outcomes for any setting, and each outcome is associated with a computable $\bm{w}$, one could use this rule to determine the relative probabilities of the two outcomes.

The fact that the weak value vectors can be associated with a qubit's worldline makes it plausible that one might use them to identify covariant and invariant quantities in a quantum circuit, as motivated in the Introduction. But it is not obvious that a complex three-vector such as $\bm{w}$ can be mathematically associated with such quantities. We now summarize how to resolve this issue, before demonstrating in Sec. \ref{sec3} that the evolution of the weak values can respect relativistic covariance.

\subsection{Covariants}\label{sec24}

In analogy to the antisymmetric relativistic field tensor from electromagnetism
\begin{equation}
    F^{\mu\nu}=\begin{pmatrix} 
    0 & E_x/c & E_y/c & E_z/c \\
    -E_x/c & 0 & B_z & -B_y \\
    -E_y/c & -B_z & 0 & B_x \\
    -E_z/c & B_y & -B_x & 0 \\
    \end{pmatrix}
\end{equation}
we propose that the 6 components of each weak value are naturally incorporated into a similar covariant tensor:
\begin{equation}
    W^{\mu\nu}=\begin{pmatrix} 
    0 & Im(\bm{w}_x) & Im(\bm{w}_y) & Im(\bm{w}_z) \\
    -Im(\bm{w}_x) & 0 & Re(\bm{w}_z) & -Re(\bm{w}_y) \\
    -Im(\bm{w}_y) & -Re(\bm{w}_z) & 0 & Re(\bm{w}_x) \\
    -Im(\bm{w}_z) & Re(\bm{w}_y) & -Re(\bm{w}_x) & 0 \\
    \end{pmatrix}.
\end{equation}

One motivation for this identification is the case of a spin-1/2 qubit, where the real weak value measurements correspond to an angular momentum vector proportional to a magnetic moment $\bm{\mu}$, associated with a magnetic field.  If a magnetic moment is boosted to an arbitrary frame, it picks up an electric dipole moment, such that the relativistic moment also takes the form of an antisymmetric tensor. 

More concretely, in analogy with the field invariants $B^2-E^2/c^2$ and $\vec{B}\cdot\vec{E}$, the invariants of such a tensor are $Re(\bm{w})^2-Im(\bm{w})^2$ and $Re(\bm{w})\cdot Im(\bm{w})$.  Recall that in the above single-qubit example, the weak values always obey $\bm{w}\cdot\bm{w}=1$, or equivalently:
\begin{eqnarray}
\label{eq:invars}
    Re(\bm{w})^2-Im(\bm{w})^2 &=& 1 \nonumber \\
    Re(\bm{w})\cdot Im(\bm{w}) &=& 0.
\end{eqnarray}
If these conditions are to be frame-invariant, then $W^{\mu\nu}$ is an appropriate covariant description of a qubit's local weak values. Further evidence for this proposal will be given below.

\section{Individual Qubits in Quantum Circuits}\label{sec3}

\subsection{Localized Weak Values}

Consider a generic N-qubit quantum circuit, where at every moment the full quantum state is generally entangled. Even for such an entangled state, a local weak value for each individual qubit can be defined as in Eqn. (\ref{eq:weak}), so long as one uses an operator ${A}$ with the tensor product form ${A}=B\otimes I$.  Here $B$ is a single qubit operator (say, one of the Pauli matrices $\sigma_j$), and $I$ is the identity operator of the appropriate dimension to act on the remaining $N-1$ qubits.

The complex-valued vector of weak values $\bm{w}$ was defined in Eq. (\ref{eq:sdef}) for a single qubit.  When considering an N-qubit system, possibly in an entangled state, the analogous expression for a single qubit of interest (say, qubit ``a'') can be generalized to
\begin{equation}
\bm{w}_a=(W[\sigma_x\otimes I],W[\sigma_y\otimes I],W[\sigma_z\otimes I]).
\end{equation}

Such a procedure will work for any individual qubit; when more than one qubit is considered, each qubit will be noted with a subscript, such as $\bm{w}_a$, $\bm{w}_b$, etc.  Of course, there are also operators that refer to multiple qubits and so are not of the form $A=\sigma_j\otimes I$, but only weak values calculated with a local operator are candidates for a localized, covariant description. 

The question at hand is whether the weak values evolve according to local conditions -- in our terminology, this would make the weak values ``dynamically local''.  Any hope for such dynamic locality would be dashed if the single-qubit weak values calculated in this way depended upon the evolution of distant qubits in the same entangled state. Another worry might be that the weak values would reveal a special ``plane of simultaneity'', reflecting the reference frame in which the state vectors are defined and preventing any possible covariance. The next subsection will show that these potential concerns do not, in fact, occur.  

\subsection{Dynamic Locality for Entangled Qubits}\label{sec32}

In the single-qubit example from \ref{sec22}, we saw that the weak value vector remained constant between preparation and measurement, so long as there was no Hamiltonian acting on it.  If there is a hidden local description underpinning the qubit of interest, $\bm{w}_a$ should also remain constant for any N-qubit Hamiltonian of the form $H_N=I\otimes H_{N-1}$, even if the other $N-1$ entangled qubits are evolving due to some Hamiltonian $H_{N-1}$.  In other words, given a Hamiltonian of this form, the weak value $W[\sigma\otimes I]$ should stay constant for any single-qubit operator $\sigma$.

To test this proposal, all one needs is to represent the unitary time-evolution operator in the form $U_N=I\otimes U_{N-1}$ and calculate the relevant weak values $W[\sigma\otimes I]$ at two different times.  Before the evolution, this yields
\begin{equation}
    W[\sigma\otimes I](t_i) =\frac{\bra{f}U_{N}(\sigma\otimes I)\ket{i}}{\braket{f|U_{N}|i}}.
\end{equation}
Meanwhile, at the final measurement time $t_f$, the same weak value is found via
\begin{equation}
    W[\sigma\otimes I](t_f) =\frac{\bra{f}(\sigma\otimes I)U_{N}\ket{i}}{\braket{f|U_{N}|i}}.
\end{equation}
Evidently\footnote{This follows because, if $A$ and $B$ are operators in the same Hilbert subspace (and the same is true for $C$ and $D$), then $[A\otimes C,B\otimes D]=AB\otimes CD - BA\otimes DC $.}, $\sigma\otimes I$ commutes with $U_N=I\otimes U_{N-1}$.  This means that the above two expressions are identical, so the weak values will remain constant along any wire, even if the qubit is part of an entangled state. 

\subsection{Single-Qubit Gates}

Having the weak values remain constant on the wires is not enough.  They will also have to respond in a reasonable way when they encounter a single-qubit gate $U_1$, even with other qubits in the entangled state.  In this case, the relevant unitary time-evolution operator would be $U_1\otimes U_{N-1}$.  Picking out an arbitrary component of the weak value vector $\bm{w}_j$ for the qubit of interest (corresponding to some Pauli matrix $\sigma_j$), the weak value before the $U_1$ gate is equal to
\begin{equation}
    W[\sigma_j\otimes I](t_i) =\frac{\bra{f}(U_1\otimes U_{N-1})(\sigma_j\otimes I)\ket{i}}{\braket{f|(U_1\otimes U_{N-1})|i}}.
\end{equation}
This will no longer be the same as the same weak value after the gate. 

However, notice that after the gate there is a different weak value component, $W[\sigma_f\otimes I]$ which would be equal to the above expression.  This component could be calculated according to
\begin{equation}
    W[\sigma_f\otimes I](t_f) =\frac{\bra{f}(\sigma_f\otimes I)(U_1\otimes U_{N-1})\ket{i}}{\braket{f|(U_1\otimes U_{N-1})|i}}.
\end{equation}
A short calculation reveals that these two equations will be identical if $U_1 \sigma_j = \sigma_f U_1$, or, equivalently, if $\sigma_f=U_1\sigma_j U_1^{-1}$.

This means that the weak value $W[\sigma_j]$, originally associated with the $\sigma_j$ direction, has now been mapped to $W[\sigma_f]$ (associated with the $\sigma_f$ direction) after passing through the single-qubit gate. And the relationship between $\sigma_j$ and $\sigma_f$ is just a rotation, exactly what the gate $U_1$ is known to do to any Bloch-sphere vector. In other words, if one knew the weak value vector $\bm{w}$ going into a single qubit gate corresponding to a known rotation, the resulting weak values could be calculated simply by subjecting the weak value vector $\bm{w}$ to precisely that same rotation.  This rotation depends only on $U_1$, not on the $U_{N-1}$ evolution affecting the other qubits. Hence the weak value evolution through the gate is independent of what is simultaneously happening to the rest of the entangled state.

\subsection{Measurements}

In the single qubit case, we saw that the weak value vector $\bm{w}$ was partially fixed in the direction $\hat{\bm{f}}$ corresponding to the measured state, even before it arrived at the measurement. Specifically, we found $\bm{w}\cdot\hat{\bm{f}}=1$, with no imaginary component in that direction. Here we will briefly show why this is always the case at any measurement, even in a multi-qubit circuit. 

Suppose the initial state vector $\ket{i}$ is subjected to any unitary evolution $U_N$, and then one qubit is measured.  The final measurement operator for the qubit of interest is $\sigma_f$, with eigenfunctions that solve $\sigma_f\ket{f_\pm}=\pm\ket{f_\pm}$.  A measurement of only that qubit (corresponding to the full operator $\sigma_f\otimes I$) will therefore result in the partially-separable final state $\ket{f}=\ket{f_\pm}\otimes\ket{\psi_{N-1}}$. 

We are interested in the weak value component $\bm{w}\cdot\hat{\bm{f}}_+=W[\sigma_f\otimes I]$ for that one qubit, just before measurement. Since the evolution $U_N$ has already occurred, this yields
\begin{equation}
    W[\sigma_f\otimes I](t_f) =\frac{\bra{f}(\sigma_f\otimes I)U_N\ket{i}}{\braket{f|U_N|i}}.
\end{equation}

However, since $\bra{f}=\bra{f_\pm}\otimes\bra{\psi_{N-1}}$, and $\bra{f_\pm}\sigma_f=\pm\bra{f_\pm}$, this simplifies to 
\begin{equation}
    W[\sigma_f\otimes I](t_f) =\frac{\pm\bra{f}U_N\ket{i}}{\braket{f|U_N|i}} = \pm 1.
\end{equation}
Therefore, if $\ket{f_+}$ is measured, then $\bm{w}\cdot\hat{\bm{f}}_+=1$.  If $\ket{f_-}$ is measured, $\bm{w}\cdot\hat{\bm{f}}_+=-1$, but this is the same as $\bm{w}\cdot\hat{\bm{f}}_-=1$.  This matches the single-qubit results from the previous section.

The above analysis can also be easily inverted to apply at preparation, where (unlike at measurement) one has additional control to choose the eigenvalue.  Again, one will find that $\bm{w}\cdot\hat{\bm{i}}=1$, where $\hat{\bm{i}}$ is the chosen preparation direction.

\subsection{Discussion}

To summarize the above results, a qubit on any ``wire'' stretching between two gates in any quantum circuit will have a weak value vector $\bm{w}$ which always remains constant (both imaginary and real parts) until it hits a gate. This is true even if it is part of an entangled state, and even if the other qubits are simultaneously passing through gates of their own. Furthermore, when one qubit passes through a single-qubit gate which implements a (Bloch sphere) rotation, the corresponding weak value will rotate in the same manner regardless of whether there is entanglement with other qubits or not. This rotation applies to both real and imaginary parts of $\bm{w}$.  These results are precisely the behavior one would expect if the weak values represented a hidden localized account of what was actually happening in any given quantum circuit.

It is important to emphasize that this analysis concerns the dynamical evolution of the weak values as they pass through a given quantum circuit, not counterfactual changes between different circuits.  Suppose circuits $C_1$ and $C_2$ start out in an identical manner, but they then use a different single-qubit gate at one point in mid-circuit.  In any all-at-once analysis of the two circuits, that difference at one point might very well lead to differences in the weak values at every point.  By comparing $C_1$ and $C_2$, basic causal reasoning then tells us that an intervention at one point (at one gate) can have effects on distant qubits, in apparent contradiction to the above analysis.

But there is no contradiction here.  The above proofs show that in both circuits $C_1$ and $C_2$ the weak values always stay constant on the circuit wires, and only dynamically change when they pass through a gate.  The apparent nonlocal influence described above (and discussed, for example, in \cite{braverman}) is not a dynamical change, but a counterfactual ``change'', only evident when comparing the complete runs of two different circuits.  Evidently, this sort of counterfactual ``influence'' must be allowable in any sort of model which can violate the Bell/CHSH inequalities and thereby agree with observed entanglement correlations.

Another related claim might be that some of the above results are simply inevitable consequences of the no-signaling theorem. This is incorrect; that theorem concerns "counterfactual changes" to local gates and measurable statistics, but says nothing about dynamical changes of hidden variables on a single circuit run. Indeed, the analysis in \cite{braverman} shows that if the weak values were measurable in a single run, then one could indeed use them to signal at a spacelike distance. The fact that the hidden, single-run weak values do not display any simultaneous reaction to what is happening at distant gates is a distinct result, unaddressed by the no-signaling theorem on its own.

\section{Dynamic Locality for Two-Qubit Gates}\label{sec4}

\subsection{A Universal Two-Qubit Gate}

When two identical qubits interact via the exchange interaction, with a Hamiltonian of the form $H_{ex}=J({\sigma}_{x}\otimes{\sigma}_{x}+{\sigma}_{y}\otimes{\sigma}_{y}+{\sigma}_{z}\otimes{\sigma}_{z})\equiv J(\bm{\sigma}\otimes\bm{\sigma}$), they evolve in a way that is equivalent to passing through a two-qubit SWAP$^\alpha$ gate.  Here $J$ is a constant which determines the coupling energy.  In the $\ket{00},\ket{01},\ket{10},\ket{11}$ basis, to within a global phase, the corresponding unitary operator for this exchange interaction is
\begin{equation}
\label{eq:Uex}
U_{ex}[\alpha] = \begin{pmatrix} 1 & 0 & 0 & 0 \\
0 & \frac{1+e^{i\pi\alpha}}{2} & \frac{1-e^{i\pi\alpha}}{2} & 0 \\
0 & \frac{1-e^{i\pi\alpha}}{2} & \frac{1+e^{i\pi\alpha}}{2} & 0 \\
0& 0&0&1
\end{pmatrix}.
\end{equation}
Here $\alpha$ is used to compute the net result of the total interaction, so it is therefore a product of the interaction strength $J$ and the total time over which the interaction persists.  In this section it will be useful to think of $J$ as being a constant over the duration of the interaction, with $\alpha$ varying to indicate this duration, scaled such that $\alpha=1$ corresponds to a full SWAP of the two qubits. 

Turning on the interaction for only half the SWAP time ($\alpha=0.5$) corresponds to a $\sqrt{\text{SWAP}}$ gate.  This gate can be used to entangle two separable qubits and two $\sqrt{\text{SWAP}}$s (combined with single qubit gates) can generate the more common CNOT gate.  It is well known that the CNOT and single-qubit gates comprise a universal set for generating any possible quantum circuit, so the same must be true for the $\sqrt{\text{SWAP}}$ gate when combined with the single-qubit gates from the previous section.

Apart from generating a universal set of gates for quantum computation, we focus on the SWAP$^\alpha$ gate for two additional reasons.  First, the exchange interaction is the most ``natural'' interaction for two charged spin-1/2 systems; such a Hamiltonian can be implemented simply by bringing two such particles to the same average location.  More importantly, when thinking of $\alpha$ as a time parameter, the SWAP$^\alpha$ represents a \textit{continuous} gate (as contrasted with a complete CNOT operation).  Since weak values can be computed without actually making a weak measurement, one can reconstruct the time-history of the weak values inside the gate simply by changing the values of $\alpha$ in the gates before and after the point at which the weak values are calculated.  Carrying out this analysis will indicate that the weak values are evolving according to a simple classical equation, even as quantum theory says that the qubits are becoming entangled. 

\subsection{Results: Dynamic Locality}

To get a sense of the behavior of the weak values in an exchange interaction, we first turn to numerical modeling.  Our code chooses a random two-qubit state $\ket{i}$ for the preparation, then evolves these two qubits with a SWAP$^\alpha$ gate (for any chosen value of $\alpha$).  The gate output is measured in a manner that corresponds to another random two-qubit state $\ket{f}$.  At this stage, there is no restriction on $\ket{i}$ or $\ket{f}$; they can each be either entangled or separable.

To parameterize the evolution of the weak values inside the SWAP$^\alpha$ gate, it is convenient to use a unitless time parameter $\tau$, ranging from $\tau=0$ at the start of the gate to $\tau=\alpha$ at the end.  (Recall that we are keeping the strength of the exchange interaction constant, so $\tau$ evolves linearly with time.)  Using the above notation, the weak value vector $\bm{w}$ for qubits $a$ and $b$ at any point in the midst of the SWAP$^\alpha$ gate can be found according to 
\begin{eqnarray}
    \bm{w_a}(\tau)&=&\frac{\bra{f} U_{ex}[\alpha-\tau] (\bm{\sigma}\otimes I)U_{ex}[\tau]\ket{i}}{\braket{f|U_{ex}[\alpha]|i}}\\ 
    \bm{w_b}(\tau)&=&\frac{\bra{f} U_{ex}[\alpha-\tau] (I\otimes \bm{\sigma})U_{ex}[\tau]\ket{i}}{\braket{f|U_{ex}[\alpha]|i}}
\end{eqnarray}

As before, these are both complex 3-vectors; complex because they are weak values and 3-vectors because of the definition of the vector of Pauli matrices $\bm{\sigma}=(\sigma_x,\sigma_y,\sigma_z)$.  Also note that unlike the many-qubit identity operators in Section \ref{sec3}, the $I$ here applies to only a single qubit.

Numerical simulation of these two complex vectors reveals that the average $\bm{w}_{avg}=(\bm{w_a}+\bm{w_b})/2$ always remains constant over time, for each of the six components (3 real and 3 imaginary).  More interestingly, each of the six components always oscillate around the corresponding component of $\bm{w}_{avg}$ as a simple harmonic oscillator.  Figure 3 shows a completely typical result for one of these components, at a series of intermediate points, for $0\le\alpha\le 2.3$.  The fit line is a sinusoidal oscillation around the average, and the results always lie on this line to within expected numerical error.

\begin{figure}[htbp]
\begin{center}
\includegraphics[width=8cm]{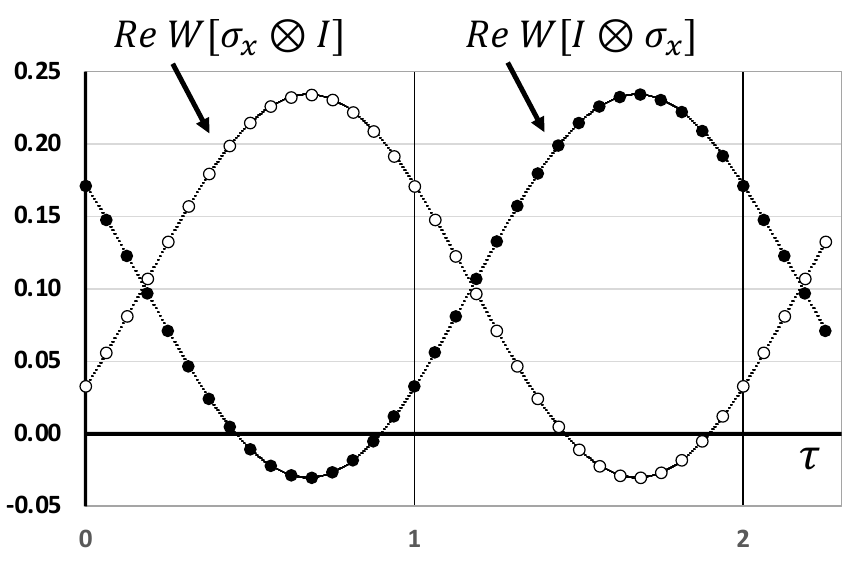}
\caption{This graph shows a typical pair of local weak value components for two qubits at various points during an exchange interaction. The essentially-perfect fit indicates that they are obeying a basic harmonic oscillator equation.  As required, the components always ``SWAP'' at $\tau=1$, and return to their original state at $\tau=2$.}
\end{center}
\end{figure}

We can infer from these results that during an exchange interaction, the weak values obey a differential equation of the precise form
\begin{eqnarray}
\label{eq:eoms}
    \frac{d^2\bm{w_a}}{d\tau^2} &=&   \frac{\pi^2}{2} (\bm{w_b}-\bm{w_a})\\
    \frac{d^2\bm{w_b}}{d\tau^2} &=&   \frac{\pi^2}{2} (\bm{w_a}-\bm{w_b}).\nonumber
\end{eqnarray}
This interaction between $\bm{w_a}$ and $\bm{w_b}$ is not non-local; in order to implement such an interaction, the qubits need to be physically brought together in space.  Physically separating them would send $J\to 0$, and the results from the previous section would then keep $\bm{w_a}$ and $\bm{w_b}$ constant.

Despite these simple classical equations, merely knowing the initial values of $\bm{w_a}$ and $\bm{w_b}$ is not sufficient to solve them. One would also need to know additional information, such as $d\bm{w_a}/d\tau$ and $d\bm{w_b}/d\tau$ at the beginning of the interaction.  Another path forward would be to use both the initial and final values of $\bm{w_a}$ and $\bm{w_b}$ to solve these equations all-at-once, as is commonly done in action extremization problems.

With these numerical results in hand (\textit{i.e.}, knowing
what to look for), it becomes possible to find general solutions for the weak value components in the $SWAP^\alpha$ gate and show that they solve Eq. (\ref{eq:eoms}).  This was checked analytically by one of us (RS), thereby providing confirmation of the above equations.  A different analytical proof, starting with Eq. (\ref{eq:weak}), can be found in Appendix \ref{aA}. 

These results were derived for a two-qubit system, but they were also checked for a multi-qubit system, where the state was fully entangled but only two qubits were interacting.  The success of this extension follows from generalizing the results in Section \ref{sec32}, parsing the system into a 2-qubit subspace and a $(N\!\!-2)$-qubit subspace, $U=U_{ex}\otimes U_{N-2}$.  As before, any entanglement is irrelevant to the dynamic locality observed in the local weak values.

\subsection{Covariant Dynamics}

In \ref{sec24}, we proposed that the weak values could be interpreted as a covariant tensor $W^{\mu\nu}$. The easiest way to show that this is consistent with Eqn. (\ref{eq:eoms}) is to present a covariant Lagrangian from which those equations can be derived.  The fact that this is possible is further evidence that $W^{\mu\nu}$ is likely the correct description.

In analogy to how the field tensor $F^{\mu\nu}$ can be built from a 4-potential $A^\mu$ and its derivatives, we can assign a 4-vector $Q^\mu(\tau)$ to each qubit's worldline (where $\tau$ is the proper time), and build the antisymmetric $W^{\mu\nu}$ from $Q$ and its canonical momentum $P$, according to
\begin{equation}
\label{eq:Wdef}
W^{\mu\nu} = Q^\mu P^\nu - P^\mu Q^\nu 
\end{equation}
A single non-interacting qubit Lagrangian $L_0$ would naturally take the form of a harmonic oscillator
\begin{equation}
\label{eq:L0}
L_0(Q^\mu,{\dot{Q}}^\mu)=\frac{1}{2}(\dot{Q}^\mu{\dot{Q}_\mu}-\omega_0^2{Q^\mu}{Q_\mu}).
\end{equation}
This notation assumes the Einstein summation convention over any repeated index, so the implied $\sum_{\mu=0}^3$ is suppressed on the right hand side.

The oscillation frequency $\omega_0$ of each qubit is taken to be a property of the qubit itself (say, for a photonic qubit, the frequency of the photon; or for a massive qubit, the Compton frequency $mc^2/\hbar$).  We will assume all qubits have identical frequencies, as in the strongest exchange interactions.

This Lagrangian is essentially that of four uncoupled harmonic oscillators, whose Euler-Lagrange equation has the general solution of
\begin{equation}
 \label{eq:Qdef}
 Q^\mu(t) = Re\left(q^\mu e^{-i\omega_0 t}\right)  
\end{equation}
At this stage, $q^\mu$ is just a constant; it is a four-vector of complex amplitudes, used to keep track of phases in the underlying $Q^\mu$ that solves a second-order-in-time differential equation. The two-qubit gate should then give a slow time variation to $q^\mu(\tau)$, changing at much slower scales than $\omega_0^{-1}$.

From the conjugate momenta $P^\mu = \dot{Q}^\mu$, a short calculation of $W^{\mu\nu}$ via (\ref{eq:Wdef}) reveals that the $exp(\pm 2\omega_0t)$ terms are symmetric under $\mu\leftrightarrow\nu$ and vanish, leaving only the constant result of $W^{\mu\nu}=Re(q^\mu) Im(q^\nu)-Im(q^\mu) Re(q^\nu)$.  Evidently, both $q^\mu$ and $W^{\mu\nu}$ describe the essential shape traced out repeatedly by the four high-frequency oscillators $Q^\mu$, without undergoing any variation themselves.  This shape will then change slowly (compared to $\omega_0^{-1}$) when the two-qubit interaction is turned on.

The simplest covariant way to couple two qubits together is with the interaction Lagrangian 
\begin{equation}
    L_{int}(Q_a,\dot{Q}_a,Q_b,\dot{Q}_b) = {\beta}(Q_a^\mu Q_{b\mu}).
\end{equation}
(The Latin indices $a,b$ are simply labels for the two different qubits; the Greek index ranges from $0$ to $3$ as usual.)  Adding this interaction in with the zero-order Lagrangians (\ref{eq:L0}) for each qubit, the total Euler-Lagrange equations then just look like coupled oscillators,
\begin{eqnarray}
    \label{eq:EL2}
    \ddot{Q}_a^\mu &=& -\omega_0^2Q_a^\mu + \beta Q_b^\mu \\
    \ddot{Q}_b^\mu &=& -\omega_0^2Q_b^\mu + \beta Q_a^\mu \nonumber
\end{eqnarray}
Coupled oscillators, of course, will ``SWAP'' after a certain amount of time, based on the coupling strength $\beta$, just like in an exchange interaction.  Pulling out the fast-$\omega_0$ oscillations, the amplitudes then vary according to the equations
\begin{eqnarray}
    \dot{q}_a^\mu &=& -i\frac{\beta}{2\omega_0} q_{b}^{\mu}\\
    \dot{q}_b^\mu &=& -i\frac{\beta}{2\omega_0} q_{a}^{\mu},
\end{eqnarray}
and the spin bivector $W_a^{\mu\nu}=Re(q_a^\mu) Im(q_a^\nu)-Im(q_a^\mu) Re(q_a^\nu)$ can be seen to oscillate in agreement with how the weak values evolve in an exchange interaction:
\begin{equation}
    \label{eq:money}
    \ddot{W}_a^{\mu\nu}=-\frac{\beta^2}{4\omega_0^2}(W_a^{\mu\nu}-W_b^{\mu\nu}).
\end{equation}
Since this equation comes from an invariant Lagrangian, it can evidently be utilized in any frame.

\section{Conclusions}\label{sec5}

To summarize the above results, we have shown that each individual qubit in any quantum circuit has well-defined ``local weak values'', that can be represented as a single covariant tensor $W^{\mu\nu}$ for each qubit. Even if the state is massively entangled, these tensors do not evolve or collapse when other qubits are measured or pass through distant circuit elements. Furthermore, the time-evolution of these tensors always obeys covariant dynamical rules. 

On the surface, there is no reason to expect to find localized, relativistic covariants in a quantum circuit. A typical circuit calculation proceeds using nonlocalized state vectors, allowing the entire state to determine the joint probability of separated outcomes. Throughout such a calculation, one assumes a preferred frame of simultaneity. 

But finding such covariants in the context of local weak values is less surprising when one considers that they can actually be measured in the laboratory, at least on average. If measurements are performed on a spin-$1/2$ system and we assume that relativity is correct, the measurement results must at least be covariant. Such measurements essentially entail measuring magnetic moments, and it is known that the values of such moments (and their associated fields) take the covariant form of an anti-symmetric tensor $F^{\mu\nu}$. This implies that the local weak value of the spin vector should also take this form. As it happens, this is exactly what has been found in the structure of $\bm{w}$, which has been naturally assigned to a covariant antisymmetric tensor $W^{\mu\nu}$.  We have also identified a special role (\ref{eq:invars}) for the two invariants encoded in $W^{\mu\nu}$ at every preparation and measurement.

It is not enough for our quantities to be expressible in covariant form. They must also evolve in a ``dynamically local'' manner. We have shown that this is always the case for a given run of a given circuit. At any point in the circuit, the local evolution is completely decoupled from what is ``simultaneously'' happening in distant gates, even if the qubits in those gates are part of the same entangled state vector. This is essentially a requirement for relativistic covariance.

Even more promising, the quantum gates we have employed are seen to always correspond to local, covariant behavior, namely a rotation for single-qubit gates, and a harmonic oscillation for $\sqrt{SWAP}$ gates, the latter satisfying the Lagrangian-compatible Eqn. (\ref{eq:money}). And because these gates form a universal set for quantum computation, these results extend to any quantum circuit with well-defined qubit worldlines, such as trapped-ion \cite{trappedion} and Rydberg atom\cite{Rydberg} implementations.

Even for qubits that are not physically localized, as in the case of which-way entanglement (with $\ket{0}$ representing a particle on one path and $\ket{1}$ a particle on another), a localized account is still available.  All that is required, as described in \cite{leifer}, is to assign a localized qubit to \textit{each} path, where $\ket{1}$ is a particle, and $\ket{0}$ is no particle (the ``vacuum state’’ from quantum field theory).  This transforms the original nonlocalized single qubit $a\ket{0}+b\ket{1}$ into \text{two} localized qubits in a larger Hilbert space, of the form $a\ket{10}+b\ket{01}$, and this would fall under the circuit framework discussed above.\footnote{This step is likely also necessary in order to make sense of anomalous (negative) weak values; as shown in \cite{wharton2018}, such values can naturally be viewed as indicating that the value has dropped below the typical (non-zero) ``vacuum state’’.} 

By analyzing local weak values we have shown that a localized physical reality, residing in spacetime rather than configuration space, seems to underlie any N-qubit entangled state. This picture is also consistent with relativistic covariance despite the instantaneous spacelike connections suggested by the orthodox state vector. Except for the single-qubit example of Section \ref{sec2}, we have not shown how to calculate the local weak values (or the probabilities) without using the full state vector as an intermediate computational tool. But this paper could be viewed as the first step towards the goal of bypassing the state vector, identifying potential covariant variables on which a direct analysis could be carried out.  The results here indicate that this could be a fruitful research program in quantum foundations, distinct from efforts that treat the state vector as a fundamental entity. The covariant weak values $W^{\mu\nu}$ introduced here offer an alternative ontology, motivating new efforts to explain them on their own terms. An example of this type of analysis can be found in section $V$ of \cite{preprint}, where the possibility was considered that weak values might be an average over an even lower-level covariant description.

In the Introduction it was noted that Barandes has argued against such a goal, using the framework of orthodox quantum theory. After analyzing various claims about the interpretation of post-selected weak values, he concluded that ``weak values do not reveal interesting physical or metaphysical properties of individual quantum systems''\cite{barandes}. One of his main arguments was that weak values are themselves only measureable over many experimental runs, and therefore are more likely ensemble (average) properties, rather than single-system properties. But as noted in the previous paragraph, this would still be consistent with our conclusions; even if the local weak values were a statistical average, looking to their covariant structure would still be a guide as to the best single-system description. His other main argument concerned post-selection fallacies, a line of reasoning which (implicitly) assumes no future-input dependence. If the post-selection includes a setting that imposes a boundary constraint on the past values of any hidden variables, the all-at-once analysis of this paper is more suitable; see \cite{pricewharton} for details on how post-selection might be viewed in a future-input dependent context.\footnote{Even in a future-input dependent model, Barandes is correct that one can not make proper inferences from the statistics of \textit{outcome}-selected subensembles (see \cite{wharton2025}), but that should not be conflated with fixing the final \textit{measurement basis}.}

It seems likely that other applications might be found for our central results. Certainly, there are many different research programs in quantum foundations that might find a use for the local weak values, from Consistent Histories \cite{griffiths} to Bohmian Mechanics \cite{sutherland2008}. More broadly, being able to utilize two Bloch-ball-like vectors to represent individual qubits in entangled states might find applications from pedagogy to circuit design. (The two vectors $Re(\bm{w})$ and $Im(\bm{w})$ can extend beyond the surface of the Bloch sphere, but would nevertheless allow one to visualize the local behavior of multi-qubit circuits.)  And, of course, these results would likely inform future research into the nature of weak values themselves, drawing further connections to previous work \cite{aharonov2015,skorobagatko2022,waegell2023}.

Even without follow-up research, our present results already challenge the widespread view that violations of Bell inequalities in quantum circuits must be due to some sort of direct influence or connection between spatially-separated parts of the circuit. Instead, the weak values reveal no appearance of any simultaneous link between distant events. The local weak values must be violating Bell's assumptions in some other way.

And indeed, weak values can be seen to instead violate the ``statistical independence'' assumption which goes into the proofs of Bell/CHSH inequalities.  Even in the single-qubit example from \ref{sec22}, the allowable weak values depend upon the future basis in which the qubit is measured.  Such future-input dependent models have been highlighted as a promising way to explain Bell inequality violations while retaining the central elements needed for relativistic invariance \cite{wharton2020}.  The fact that analysis of the local weak values leads to precisely this sort of model could be seen as further evidence that this is a plausible way to better describe what might be really happening between measurements.  If successful, it would also support the ``all-at-once'' view of spacetime, as opposed to any perspective where time is thought to have some ``flow'' analog in the forward time direction.

The above results comprise several strands of evidence that relativistically-local quantities can be assigned to every point of a quantum circuit.  Already, this motivates a natural extension of quantum theory to include these local weak values as an additional element of reality.  But for many researchers in quantum foundations, the goal is not merely to extend quantum mechanics, but to reformulate the theory in a way that provides new explanatory power.  The fact that the weak values are seen to evolve locally, even in massively entangled states, is evidence that this approach could lead to such a reformulation. Dressel has noted that if a model ``could really mimic the detailed functional structure of the weak value, then it would also be able to simulate other features that are normally considered to be quantum mechanical’’ \cite{dressel2015}.  If this could indeed be achieved, at the mere expense of using final measurement settings as an explanatory resource, then a concerted push to find such a reformulation should arguably be a central priority for modern research in quantum foundations.

\section*{Acknowledgements}

The authors gratefully thank Nathan Argaman, Jacob Barandes, Eliahu Cohen, and Travis Norsen for helpful feedback.

\begin{appendices}

\section{}\label{aA}

Here we demonstrate that Eq. (\ref{eq:eoms}) follows from (\ref{eq:weak}) and (\ref{eq:Uex}).  We need only prove that
\begin{equation}
\label{eq:prove}
    \frac{d^2 w_a}{d \tau ^2} = \frac{\pi^2}{2}(w_b - w_a),
\end{equation}
where $w_a$ and $w_b$ are any arbitrary component of the full vectors $\bm{w}_a$ and $\bm{w}_b$.  Also, the second equation in (\ref{eq:eoms}) follows from interchanging the two qubits.  

Consider the limit definition of the second derivative:
\begin{equation}
\label{eq:p1}
    \frac{d^2 w_a}{d \tau ^2} = \lim_{\Delta \tau \to 0} \frac{w_a(\tau + \Delta \tau) - 2w_a(\tau) + w_a(\tau - \Delta \tau)}{\Delta \tau^2}
\end{equation}
The exchange interaction for a generic $SWAP^\alpha$ gate runs to a time $\tau=\alpha$, so these weak value terms can be found from:
\begin{eqnarray}
     w_a(\tau + \Delta \tau) &=& \frac{\langle f|U_{ex}(\alpha - \tau)U_{ex}^\dagger(\Delta \tau) (\sigma \otimes I) U_{ex}(\Delta \tau)U_{ex}(\tau) | i \rangle}{\langle f|U_{ex}(\alpha)|i \rangle}\nonumber \\
     w_a(\tau) &=&  \frac{\langle f|U_{ex}(\alpha - \tau)(\sigma \otimes I) U_{ex}(\tau)| i \rangle}{\langle f|U_{ex}(\alpha)|i \rangle} \nonumber\\
     w_a(\tau - \Delta \tau) &=&  \frac{\langle f|U_{ex}(\alpha - \tau) U_{ex}(\Delta \tau) (\sigma \otimes I) U_{ex}^\dagger(\Delta \tau)U_{ex}(\tau)| i \rangle}{\langle f|U_{ex}(\alpha)|i \rangle}  \nonumber
\end{eqnarray}
where $U_{ex}^\dagger(\Delta \tau) = U_{ex}(-\Delta \tau) = U_{ex}^{-1}(\Delta \tau)$, and $\sigma$ is any of the Pauli matrices $\sigma_x $, $\sigma_y $, $\sigma_z $ (corresponding to the component of interest of $\bm{w}_a$ and $\bm{w}_b$).
Then, setting $\langle f'| = \langle f|U_{ex}(\alpha - \tau)$ and $U_{ex} (\tau)|i\rangle = |i'\rangle$, Eq. (\ref{eq:p1}) becomes:
\begin{equation}
\label{eq:p2}
     \frac{d^2 w_a}{d \tau ^2} = \lim_{\Delta \tau \to 0} \frac{\langle f'|U_{ex}^\dagger(\Delta \tau) (\sigma \otimes I) U_{ex}(\Delta \tau) - 2(\sigma \otimes I) + U_{ex}(\Delta \tau) (\sigma \otimes I) U_{ex}^\dagger(\Delta \tau) |i'\rangle}{{\Delta \tau^2}\langle f|U_{ex}(\alpha)|i \rangle}.
\end{equation}

Expanding $U_{ex}(\Delta \tau)$ to second order in $\Delta \tau$, it happens that
\begin{align*}
    U_{ex}(\Delta \tau) \approx  I + \left( i \frac{\pi \Delta \tau}{2} - \frac{\pi^2 \Delta \tau^2}{4} \right)  \Big ( \frac{I - \sigma_x \otimes \sigma_x -\sigma_y \otimes \sigma_y - \sigma_z \otimes \sigma_z}{2} \Big ).
\end{align*}
Defining $M=(I-\bm{\sigma}\otimes\bm{\sigma})/2$, and a complex constant $a$, this can be expressed as $U_{ex}=I+aM$, dropping terms of order $(\Delta\tau)^3$ and higher.

With these definitions, (\ref{eq:p2}) then becomes:
\begin{equation}
\frac{d^2 w_a}{d \tau ^2} = \frac{\langle f'| 2 aa^*  M(\sigma \otimes I) M + (a + a^*) ( M (\sigma \otimes I) + (\sigma \otimes I) M )|i'\rangle } {{\Delta \tau^2}\langle f'|i' \rangle}.
\end{equation}
This expression greatly simplifies, because of the easily-checked relationships
\begin{eqnarray}
    M(\sigma \otimes I) M &=& 0\\
    a + a^* &=& -\frac{\pi^2 \Delta \tau^2}{2}\\
    M (\sigma \otimes I) + (\sigma \otimes I) M &=& (\sigma \otimes I) - (I \otimes \sigma).
\end{eqnarray}
Again, this final equation is correct for any given component $\sigma$ of the Pauli vector $\bm{\sigma}$.

Finally, using the fact that $w_b(\tau) =\bra{f'} I \otimes \sigma\ket{i'}/\braket{f'|i'}$, we find the desired equation
\begin{align*}
    \frac{d^2 w_a}{d \tau ^2} =  \frac{\pi^2 \Delta \tau^2 \langle f'| I \otimes \sigma - \sigma \otimes I |i'\rangle}{2\Delta \tau^2 \langle f'|i' \rangle}  = \frac{\pi^2}{2} (w_b - w_a).
\end{align*}

\end{appendices}

\bibliography{References}

\end{document}